# Does strange meson condensation reduce the moment of inertia of massive proto neutron stars? insights at $S = 1$ and $Y_L = 0.4$

Cheng-Jian Nie[1], and Xian-Feng Zhao[2†]

[1]School of Intelligent Connected and New Energy Vehicles, Geely University of China, Chengdu, 641423, China
[2]School of Sciences, Southwest Petroleum University, Chengdu 610500, China

**Abstract.** We investigate the effects of strange mesons ($\sigma^*$ and $\phi$) on the structural and rotational properties of massive proto neutron stars (PNSs) within the relativistic mean-field (RMF) framework. Using the TW99 parametrization with entropy per baryon $S$=1 and lepton fraction $Y_L$=0.4, we solve the TOV equations and compute the moment of inertia via the Hartle–Thorne approximation. Among eight parametrizations, TW99 is chosen as it gives the largest maximum mass. Our results show that strange mesons soften the equation of state at high densities, reducing both the maximum mass and radius. The moment of inertia peaks at a lower central density than the maximum mass, and strange mesons shift this peak toward higher densities while slightly lowering its value. The relative change in I is negligible for PNSs below 2.1 $M_\odot$, but decreases to about -0.6% at 2.7 $M_\odot$, indicating that strangeness effects are most relevant for the most massive stars. Our findings provide useful constraints on the rotational evolution of massive PNSs and their potential gravitational-wave signatures.

## 1 Introduction

Neutron stars (NSs) are unique natural laboratories for exploring dense nuclear matter under extreme conditions [1, 2,3,4]. The equation of state (EoS) of supra-nuclear matter remains a central challenge in astrophysics and high-energy physics, particularly at densities where exotic degrees of freedom such as hyperons or strange mesons are expected to appear .

The moment of inertia is a fundamental physical quantity for understanding the rotational properties of NSs. Starting from general relativity, Hartle derived the formula for calculating the moment of inertia of a slowly rotating, spherically symmetric NS [5,6]. This formula establishes the complex relation between the moment of inertia and the mass-radius profile of the star. Bejger et al. (2005) studied the moment of inertia of the NS PSR J0737-3039A (with mass $M$=1.338 $M_\odot$) using five different EoSs, finding a range of $0.98\times10^{45}$ to $1.72\times10^{45}$ g cm$^2$ [7]. Later, Raithel et al. (2016) refined this value to approximately $1.3\times10^{45}$ g cm$^2$ using 41 different EoSs [8].

The discovery of massive NSs has significantly advanced our understanding of dense matter. Notable examples include PSR J1614-2230 with a mass of 1.93 ±0.07 $M_\odot$ (Demorest et al., 2010; Fonseca et al., 2016) [9,10], PSR J0348+0432 with 2.01 ±0.04 $M_\odot$ (Antoniadis et al., 2013) [11], and PSR J0740+6620 with $2.14^{+0.10}_{-0.09}$ $M_\odot$ (Cromartie et al., 2020) [12]. These observations have effectively excluded many overly soft EoS models. Among the most extreme rotators, the black widow pulsar PSR J0952-0607, located in the Milky Way, spins at a remarkable frequency of 709 Hz, making it one of the fastest-spinning NSs identified to date (Romani et al., 2022) [13]. More recently, its mass has been determined to be 2.35 ±0.11 $M_\odot$ (Romani et al., 2026) [14], placing it among the heaviest confirmed NSs. This exceptionally massive object offers a unique opportunity to probe the properties of nuclear matter under extreme densities.

The relativistic mean-field (RMF) theory provides a robust framework for calculating the properties of NS matter. In its original formulation, only nucleons interacting via $\sigma$, $\omega$, and $\rho$ mesons were considered [15] . Later, the inclusion of the baryon octet and additional mesonic degrees of freedom became necessary to describe hyperon interactions [16]. The $f_0(975)$ (denoted as $\sigma^*$) and $\phi(1020)$ (denoted as $\phi$) mesons were introduced to describe the interactions between hyperons, while the $\delta$ meson was included to account for isospin asymmetry effects.

Proto-neutron stars (PNSs), formed immediately after core-collapse supernovae, are extremely hot and lepton-rich objects with trapped neutrinos [17,18]. Unlike cold, catalyzed NSs, PNSs require a finite-temperature EoS that depends on entropy per baryon and lepton fraction [19,20]. Recent studies have shown that the macroscopic properties of PNSs are significantly affected by thermal effects

*Send offprint requests to*: email:†zhaopioneer.student@sina.com

and lepton fraction. For example, higher entropy increases the maximum mass and flattens the mass-radius curves, while higher lepton fraction reduces the maximum mass and increases the canonical radius [21,22].

Despite these studies, a systematic investigation of the effects of strange mesons ( $\sigma^*$ and $\phi$) on the rotational properties of massive PNSs under fixed entropy per baryon ($S$=1) and lepton fraction ($Y_L$=0.4)—conditions relevant to the early evolutionary stage of NSs—remains limited. In particular, the behavior of the moment of inertia near its maximum and its shift induced by strange mesons have not been fully explored.

In this work, we systematically investigate the effects of $\sigma^*$ and $\phi$ mesons on the structural and rotational properties of massive PNSs within the RMF framework [23].

## 2 Theoretical Framework for PNSs

The Lagrangian density of a PNS at finite temperature [24,16] is given by

$$\begin{aligned}\mathcal{L} = &\sum_B \overline{\Psi}_B \left(i\gamma_\mu\partial^\mu - m_B + g_{\sigma B}\sigma + g_{\sigma^* B}\sigma^*\right.\\ &\left.-g_{\omega B}\gamma_\mu\omega^\mu - g_{\phi B}\gamma_\mu\phi^\mu - g_{\rho B}\gamma_\mu\tau_3\rho^\mu\right)\Psi_B\\ &-\frac{1}{2}m_\sigma^2\sigma^2 - \frac{1}{3}g_2\sigma^3 - \frac{1}{4}g_3\sigma^4\\ &+\frac{1}{2}m_\omega^2\omega_\mu\omega^\mu + \frac{1}{2}m_\rho^2\rho_\mu\rho^\mu - \frac{1}{2}m_{\sigma^*}^2\sigma^{*2} + \frac{1}{2}m_\phi^2\phi_\mu\phi^\mu\\ &+\sum_{\lambda=e,\mu}\overline{\Psi}_\lambda\left(i\gamma_\mu\partial^\mu - m_\lambda\right)\Psi_\lambda. \qquad (1)\end{aligned}$$

In the RMF approximation, the meson fields are replaced by their classical expectation values.

The total energy density $\varepsilon$ and pressure $p$ of the system are [19,20]

$$\begin{aligned}\varepsilon = &\frac{1}{2}m_\sigma^2\sigma^2 + \frac{1}{2}m_{\sigma^*}^2\sigma^{*2} + \frac{1}{3}g_2\sigma^3 + \frac{1}{4}g_3\sigma^4\\ &+\frac{1}{2}m_\omega^2\omega_0^2 + \frac{1}{2}m_\phi^2\phi^2 + \frac{1}{2}m_\rho^2\rho_{03}^2\\ &+\sum_B\frac{2J_B+1}{2\pi^2}\int_0^\infty \kappa^2 n_B(\kappa)\,d\kappa\,\sqrt{\kappa^2+m_B^{*2}}, \qquad (2)\end{aligned}$$

$$\begin{aligned}p = &-\frac{1}{2}m_\sigma^2\sigma^2 - \frac{1}{2}m_{\sigma^*}^2\sigma^{*2} - \frac{1}{3}g_2\sigma^3 - \frac{1}{4}g_3\sigma^4\\ &+\frac{1}{2}m_\omega^2\omega_0^2 + \frac{1}{2}m_\phi^2\phi^2 + \frac{1}{2}m_\rho^2\rho_{03}^2\\ &+\frac{1}{3}\sum_B\frac{2J_B+1}{2\pi^2}\int_0^\infty \frac{\kappa^4}{\sqrt{\kappa^2+m_B^{*2}}}\,n_B(\kappa)\,d\kappa. \qquad (3)\end{aligned}$$

Here, $n_B(\kappa)$ is the Fermi-Dirac distribution function for baryon $B$ at temperature $T$,

$$n_B(\kappa) = \frac{1}{1+\exp\left[\left(\sqrt{\kappa^2+m_B^{*2}}-\mu_B\right)/T\right]}. \qquad (4)$$

The effective baryon mass $m_B^* = m_B - g_{\sigma B}\sigma - g_{\sigma^* B}\sigma^*$ includes the attractive scalar interactions.

The structure of a PNS is determined by solving the Tolman-Oppenheimer-Volkoff (TOV) equations [25,26]

$$\frac{dp}{dr} = -\frac{(p+\varepsilon)(M+4\pi r^3 p)}{r(r-2M)}, \qquad (5)$$

$$\frac{dM}{dr} = 4\pi r^2\varepsilon, \qquad (6)$$

where $M(r)$ is the enclosed gravitational mass at radius $r$.

For a slowly rotating PNS, the moment of inertia $I$ is given by [5,6]

$$I = \frac{8\pi}{3}\int_0^R dr\, r^4\,\frac{\varepsilon+p}{\sqrt{1-2M(r)/r}}\,\frac{\bar{\omega}(r)}{\Omega}\,e^{-\nu(r)}, \qquad (7)$$

where $\Omega$ is the angular velocity of the star, and $\bar{\omega}(r) = \Omega - \omega(r)$, with $\omega(r)$ being the local angular velocity of the inertial frame. The metric function $\nu(r)$ satisfies

$$\frac{d\nu(r)}{dr} = -\frac{1}{\varepsilon+p}\frac{dp}{dr}. \qquad (8)$$

The angular velocity function $\bar{\omega}(r)$ obeys the differential equation

$$-\frac{1}{r^4}\frac{d}{dr}\left(r^4 j(r)\frac{d\bar{\omega}}{dr}\right) + \frac{4}{r}\frac{dj}{dr}\bar{\omega} = 0, \qquad (9)$$

with

$$j(r) = e^{-\nu(r)}\sqrt{1-2M(r)/r}. \qquad (10)$$

The boundary conditions are

$$\left.\frac{d\bar{\omega}}{dr}\right|_{r=0} = 0, \qquad (11)$$

$$\nu(\infty) = 0, \qquad (12)$$

$$\bar{\omega}(R) = \Omega - \frac{R}{3}\left.\frac{d\bar{\omega}}{dr}\right|_{r=R}. \qquad (13)$$

## 3 Parameters

To model PNS matter, we adopt eight distinct sets of nucleonic coupling parameters: DD-ME1 [27], NL1 [28], FSU2R [29], FSUGold [30], GL85 [15], GL97 [24], GM1 [31], and TW99 [27]. Among these, we select the parametrization that yields the highest gravitational mass for describing massive PNSs, as it provides the stiffest EoS required to support such objects. During this selection, the contributions from the strange mesons $\sigma^*$ and $\phi$ are excluded, and we fix the entropy per baryon at $S = 1$ and the lepton fraction at $Y_L = 0.4$.

For low baryon densities in the range $4.73\times 10^{-15}$ fm$^{-3}$ $< \rho <$ $8.907\times 10^{-3}$ fm$^{-3}$, we employ the BPS EoS [24]. Above this threshold, i.e., for $\rho > 8.907 \times 10^{-3}$ fm$^{-3}$, we use the EoS derived from the chosen nuclear parametrization for PNS matter.

The hyperon coupling constants $g_{\sigma h}$, $g_{\omega h}$, and $g_{\rho h}$ (where $h$ denotes $\Lambda$, $\Sigma$, or $\Xi$) are expressed relative to their nucleonic counterparts via the ratios $x_{\sigma h} = g_{\sigma h}/g_{\sigma}$, $x_{\omega h} = g_{\omega h}/g_{\omega}$, and $x_{\rho h} = g_{\rho h}/g_{\rho}$. The values of $x_{\rho h}$ are fixed according to the SU(6) quark-model symmetry [32, 33]. Since previous studies have shown that the PNS mass increases with both $x_{\sigma h}$ and $x_{\omega h}$ [34], we choose a relatively large value $x_{\omega h}$=0.9 to maximize the mass. The scalar couplings $x_{\sigma h}$ are then determined by the hyperon potential depths in saturated nuclear matter, $U_h^{(N)}$, using the relation [24]:

$$U_h^{(N)} = m_n\left(\frac{m_n^*}{m_n} - 1\right)x_{\sigma h} + \left(\frac{g_\omega}{m_\omega}\right)^2 \rho_0 x_{\omega h}, \qquad (14)$$

with the potential depths adopted as $U_\Lambda^{(N)}$ = -30 MeV [33, 35,36], $U_\Sigma^{(N)}$ = 30 MeV [33,35,36,37], and $U_\Xi^{(N)}$ = -14 MeV [38].

The hyperon–hyperon interactions, mediated by the $\sigma^*$ and $\phi$ mesons, are parameterized following Ref. [16]:

$$g_{\phi\Xi} = 2g_{\phi\Lambda} = 2g_{\phi\Sigma} = -\frac{2\sqrt{2}}{3}g_\omega, \qquad (15)$$

$$g_{\sigma^*\Lambda}/g_\sigma = g_{\sigma^*\Sigma}/g_\sigma = 0.69, \qquad (16)$$

$$g_{\sigma^*\Xi}/g_\sigma = 1.25. \qquad (17)$$

Figure 1 displays the gravitational mass as a function of the central baryon density for PNSs with entropy per baryon $S$=1 and lepton fraction $Y_L$=0.4 , calculated using eight representative nucleonic parametrizations within the RMF framework. Among these models, TW99 yields the highest maximum gravitational mass and is therefore adopted as the baseline EoS for describing massive PNSs in the present study, with the contribution of strange mesons ( $\sigma^*$ and $\phi$ ) switched off in this particular figure.

A notable feature revealed in this figure is the relative offset between the locations of the maximum mass ( $M_{\max}$ ) and the maximum moment of inertia ( $I_{\max}$ ) for the TW99 parametrization (also see Table 1). Specifically, $I_{\max}$ occurs at a significantly lower central density ( $\rho_c^{I_{\max}}$ ) compared to the density at which $M_{\max}$ is reached ( $\rho_c^{M_{\max}}$ ), with a density separation of $\Delta\rho_c = \rho_c^{M_{\max}}$ - $\rho_c^{I_{\max}}$ = 0.2524 fm$^{-3}$ . This offset is physically expected and originates from the distinct sensitivity of the moment of inertia to the stellar radius. Unlike the gravitational mass, which is predominantly governed by the high-density core via the TOV equations, the moment of inertia scales approximately as $I \sim MR^2$ and is thus strongly influenced by the radial distribution in the intermediate-density region. As the central density increases, the rapid contraction of the stellar radius eventually suppresses the growth of $I$ , causing its maximum to appear well before the TOV limit is attained.

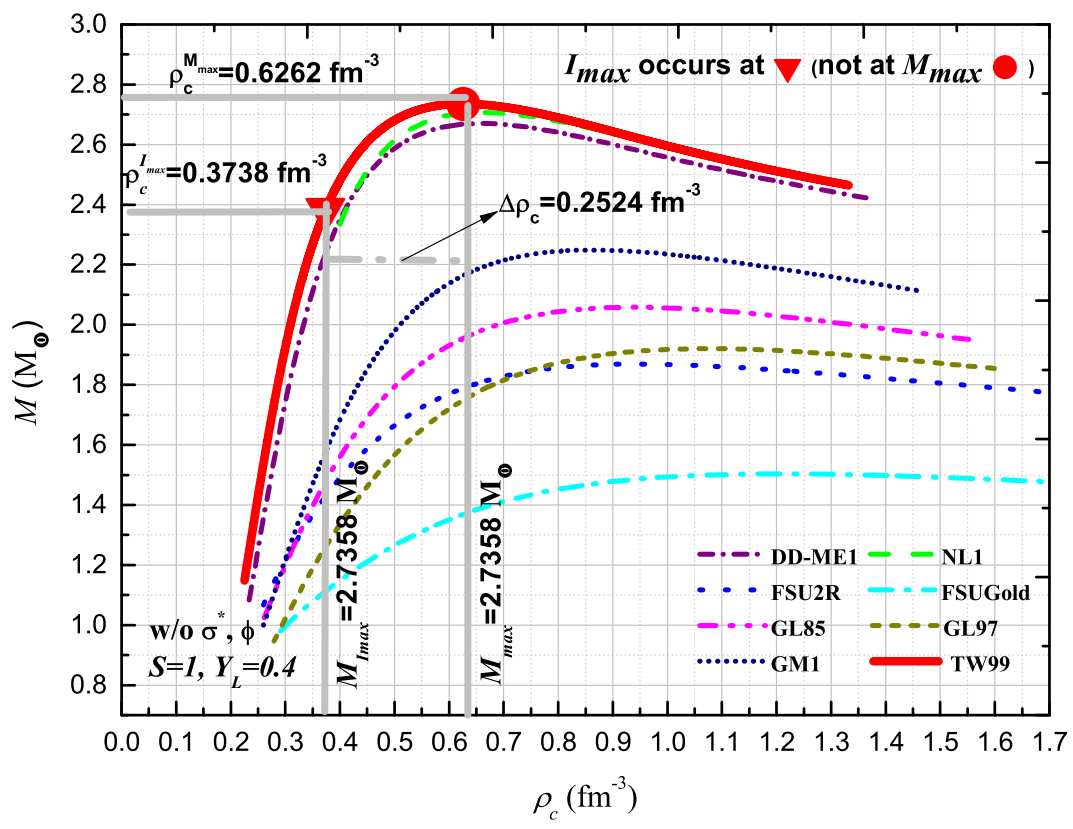

**Fig. 1.** Gravitational mass as a function of central baryon density for PNSs with entropy per baryon $S$=1 and lepton fraction $Y_L$=0.4 . Calculations are performed within the RMF model using eight different nucleonic parametrizations: DD-ME1, NL1, FSU2R, FSUGold, GL85, GL97, GM1, and TW99. Strange mesons $\sigma^*$ and $\phi$ are not included in this calculation. Among these parametrizations, TW99 yields the largest maximum gravitational mass and is therefore selected to describe the properties of massive PNSs in the present work, with its mass–density curve highlighted as a thick red solid line. The positions of the maximum mass and the maximum moment of inertia for the TW99 parametrization are indicated by symbols, with the corresponding central baryon densities labeled. Notably, the maximum moment of inertia occurs at a lower central density than the maximum mass, with a density difference of $\Delta\rho_c$ = 0.2524 fm$^{-3}$ between these two characteristic points.

This separation carries important implications for the rotational evolution of massive PNSs. Since $I_{\max}$ is located on the left (lower-density) branch of the mass–density sequence, the moment of inertia begins to decrease while the star can still support additional mass through the stiffening of the EoS. Consequently, the density interval between $I_{\max}$ and $M_{\max}$ marks a transitional regime where the PNS experiences a reduction in rotational inertia while its mass continues to grow. This behavior may significantly affect the spin-down rate and the angular momentum budget during the early cooling phase, and could leave observable imprints on the gravitational-wave signatures of future supernova events.

## 4 Equation of State

Figure 2 presents the pressure $p$ as a function of the baryon density $\rho$ for PNSs with the TW99 parametrization, entropy per baryon $S$=1 , and lepton fraction $Y_L$=0.4 , with and without the inclusion of strange mesons ( $\sigma^*$ and $\phi$ ). The red solid curve shows the pressure–density relation without strange mesons, while the green dashed curve shows the corresponding relation when strange mesons are included. When the density approaches the value corresponding to the maximum moment of inertia, a notable softening of the EoS is observed due to the appearance of strange mesons, leading to systematically lower pressures compared to the case without them. This softening originates from the attractive interactions mediated by the $\sigma^*$ and $\phi$ mesons among hyperons, which effectively reduce the internal pressure required to support the star against gravitational collapse.

**Table 1.** Structural properties of PNSs at the maximum mass $M_{\max}$ and at the maximum moment of inertia $I_{\max}$, calculated with the TW99 parametrization, entropy per baryon $S$=1, and lepton fraction $Y_L$=0.4. For each characteristic point, the corresponding radius $R$, central baryon density $\rho_c$, central energy density $\varepsilon_c$, central pressure $p_c$, and moment of inertia $I$ are listed. Results are shown both without and with the inclusion of strange mesons ($\sigma^*$ and $\phi$), along with their changes.

| **$M_{\max}$ and the associated physical quantities** | | | |
|---|---|---|---|
| Parameter | $M_{max}$ | $R^{M_{max}}$ | $\rho_c^{M_{max}}$ |
| unit | $M_\odot$ | km | fm$^{-3}$ |
| w/o $\sigma^*$, $\phi$ | 2.7358 | 13.799 | 0.6262 |
| w/ $\sigma^*$, $\phi$ | 2.7339 | 13.786 | 0.6271 |
| rate of change | -0.07% | -0.09% | 0.14% |
| Parameter | $\varepsilon_c^{M_{max}}$ | $p_c^{M_{max}}$ | $I^{M_{max}}$ |
| unit | $10^{15}$ g cm$^{-3}$ | $10^{35}$ dyne cm$^{-2}$ | $10^{45}$ g cm$^2$ |
| w/o $\sigma^*$, $\phi$ | 1.4829 | 5.8839 | 2.5543 |
| w/ $\sigma^*$, $\phi$ | 1.4855 | 5.9212 | 2.5479 |
| rate of change | 0.18% | 0.63% | -0.25% |
| **$I_{\max}$ and the associated physical quantities** | | | |
| Parameter | $M^{I_{max}}$ | $R^{I_{max}}$ | $\rho_c^{I_{max}}$ |
| unit | $M_\odot$ | km | fm$^{-3}$ |
| w/o $\sigma^*$, $\phi$ | 2.3830 | 15.149 | 0.3738 |
| w/ $\sigma^*$, $\phi$ | 2.3834 | 15.148 | 0.3742 |
| rate of change | 0.02% | -0.01% | 0.11% |
| Parameter | $\varepsilon_c^{I_{max}}$ | $p_c^{I_{max}}$ | $I_{max}$ |
| unit | $10^{15}$ g cm$^{-3}$ | $10^{35}$ dyne cm$^{-2}$ | $10^{45}$ g cm$^2$ |
| w/o $\sigma^*$, $\phi$ | 0.7395 | 1.6900 | 3.8537 |
| w/ $\sigma^*$, $\phi$ | 0.7404 | 1.6923 | 3.8489 |
| rate of change | 0.12% | 0.14% | -0.12% |

An important feature highlighted in this figure is the marked shift in the location of the maximum moment of inertia ( $I_{\max}$ ) when strange mesons are included. The red inverted triangle indicates the density and pressure at which $I_{\max}$ occurs in the absence of strange mesons, whereas the green five-pointed star marks the corresponding point with strange mesons. The inclusion of strange mesons pushes the $I_{\max}$ point toward higher densities and pressures, reflecting the modified radial distribution and the reduced stiffness of the EoS. This shift is consistent with our earlier observation in Figure 1, where the maximum moment of inertia was found to be located at a lower density than the maximum mass. The pressure increase in the high-density regime leads to a more compact star for a given central density, which in turn shifts the density threshold at which the moment of inertia reaches its peak. This sensitivity of $I_{\max}$ to the pressure behavior in the intermediate- and high-density regions underscores the crucial role of the EoS, particularly the strange meson sector, in shaping the rotational properties of massive PNSs.

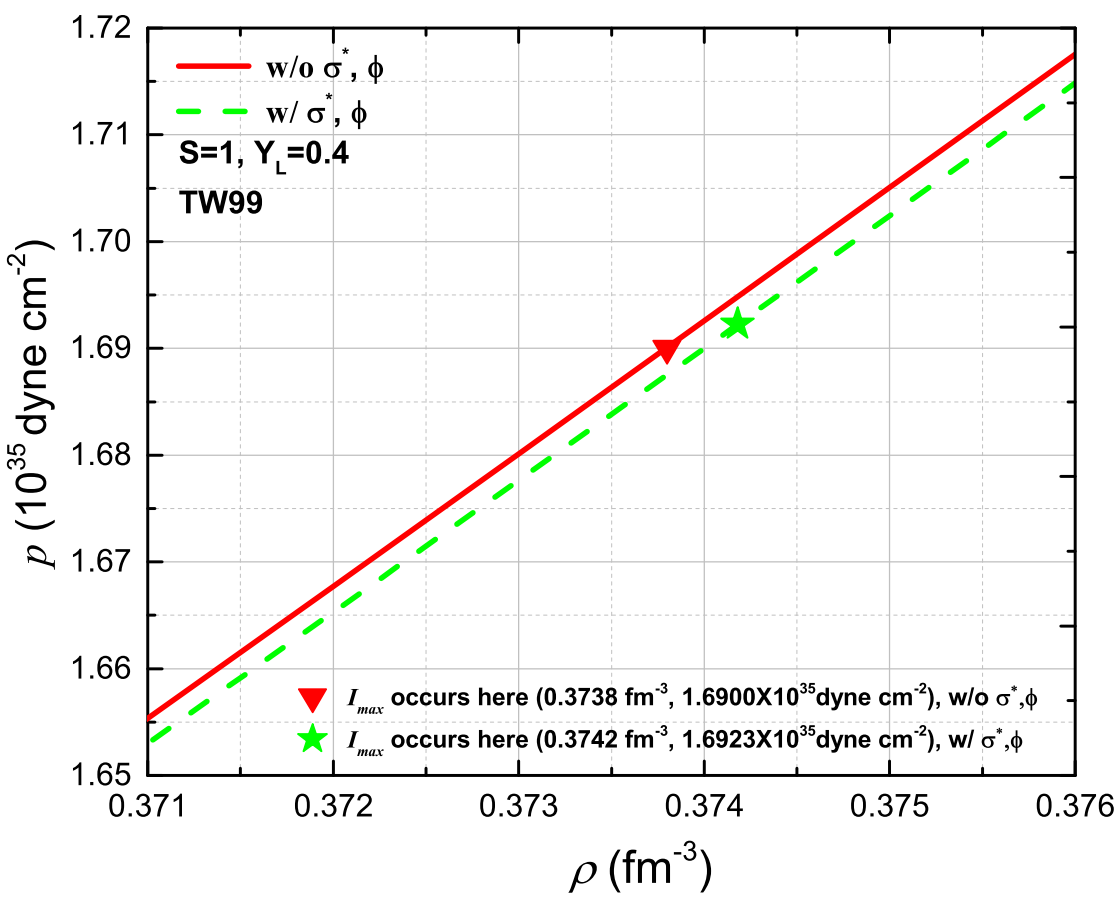

**Fig. 2.** Pressure as a function of baryon density for PNSs with the TW99 parametrization, entropy per baryon $S$=1 , and lepton fraction $Y_L$=0.4. The red solid curve represents the case without strange mesons ( $\sigma^*$ and $\phi$ ), while the green dashed curve corresponds to the case with strange mesons included. The red inverted triangle marks the point where the moment of inertia reaches its maximum in the absence of strange mesons, and the green five-pointed star indicates the corresponding point when strange mesons are included. Both markers highlight the pressure and baryon density at which the maximum moment of inertia occurs for each scenario..

Figure 3 displays the relative percentage change of pressure $\delta p$, defined as

$$\delta p = \frac{\Delta p}{p_{without}} = \frac{p_{with} - p_{without}}{p_{without}} \times 100\%, \tag{18}$$

as a function of the baryon density, for PNSs with the TW99 parametrization, entropy per baryon $S$=1 , and lepton fraction $Y_L$=0.4 . Here, $p_{without}$ and $p_{with}$ are the pressures without and with strange mesons, respectively. Positive values indicate that the pressure increases when strange mesons ( $\sigma^*$ and $\phi$ ) are included, while negative values indicate a decrease.

The curve exhibits several distinct regions that reflect the density-dependent influence of strange mesons on the EoS. At densities below approximately 0.3 fm$^{-3}$ , the relative change remains close to zero, indicating that strange mesons are not yet active and the EoS is essentially unaffected by strangeness. As the density increases into the intermediate region, the relative change becomes increasingly negative, reaching a minimum of approximately 0.5 fm$^{-3}$. This softening is attributed to the attractive interactions mediated by the $\sigma^*$ and $\phi$ mesons, which reduce the effective pressure for a given density and thereby facilitate the formation of a more compact stellar configuration.

The most notable feature in Figure 3 is the behavior near the density corresponding to the maximum moment of inertia (around $\rho$=0.37 fm$^{-3}$). In this region, the pressure reduction reaches its maximum magnitude, signifying that the rotational properties are most sensitive to the presence of strange mesons precisely where the moment of inertia peaks. This observation is consistent with our earlier findings in Figures 1 and 2, where the $I_{\max}$ point was shown to shift toward higher densities when strange mesons are included. The enhanced softening at these densities directly affects the radial distribution and, consequently, the moment of inertia, reinforcing the conclusion that the intermediate-density regime plays a pivotal role in determining the rotational characteristics of massive PNSs.

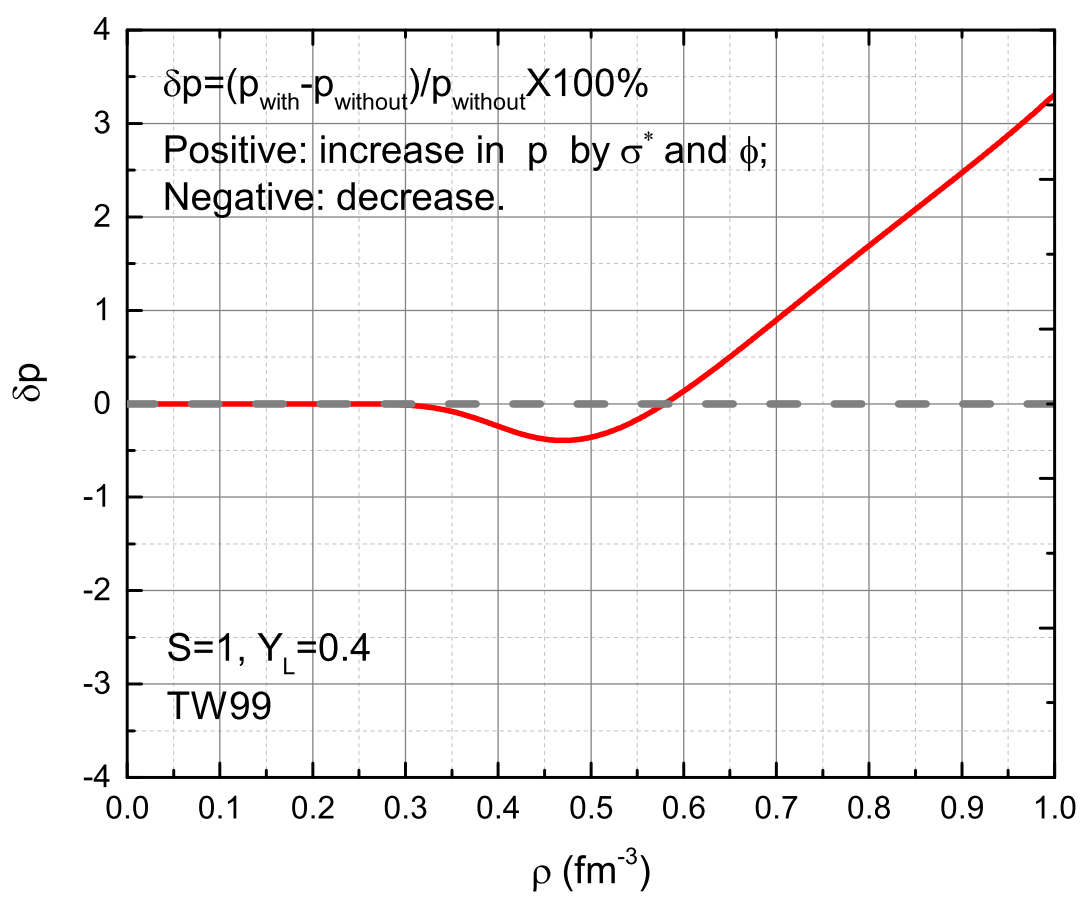

**Fig. 3.** Relative percentage change of pressure as a function of baryon density. Positive values indicate that the pressure increases when strange mesons ($\sigma^*$ and $\phi$) are included, while negative values indicate a decrease. The results are obtained for PNSs with entropy per baryon $S$=1, lepton fraction $Y_L$=0.4, and the nucleonic parametrization TW99.

At even higher densities (arund 0.6 fm$^{-3}$) approaching the maximum mass, the relative pressure change shows a tendency to recover, at which point the relative pressure change becomes positive. This behavior suggests that while strange mesons continue to soften the EOS, the effect becomes less pronounced as the core approaches its maximum gravitational mass. This recovery may be attributed to the increasing contribution of repulsive interactions or to the saturation of hyperon fractions, which limits the further softening of the EoS.

Overall, the density-dependent pressure modification quantified in Fig. 3 provides a clear physical explanation for the observed shifts in both the maximum mass and the moment of inertia. The maximum softening occurs in the density regime where $I_{\max}$ is located, confirming that the presence of strange mesons influences the rotational properties primarily through the softening of the EoS in the intermediate-density region. This sensitivity underscores the importance of accurately constraining the strange meson sector in the RMF model, particularly for predicting the rotational evolution and observable signatures of massive PNSs.

## 5 Moment of Inertia

Figure 4 displays the moment of inertia $I$ as a function of the central baryon density $\rho_c$ for PNSs with the TW99 parametrization, entropy per baryon $S$=1, and lepton fraction $Y_L$=0.4. The left panel presents the overall behavior of $I$ over a broad density range, while the right panel provides a magnified view in the vicinity of the maximum moment of inertia ( $I_{\max}$ ) to highlight the detailed structure near the peak. The red solid and green dashed curves correspond to the cases without and with strange mesons ( $\sigma^*$ and $\phi$ ), respectively.

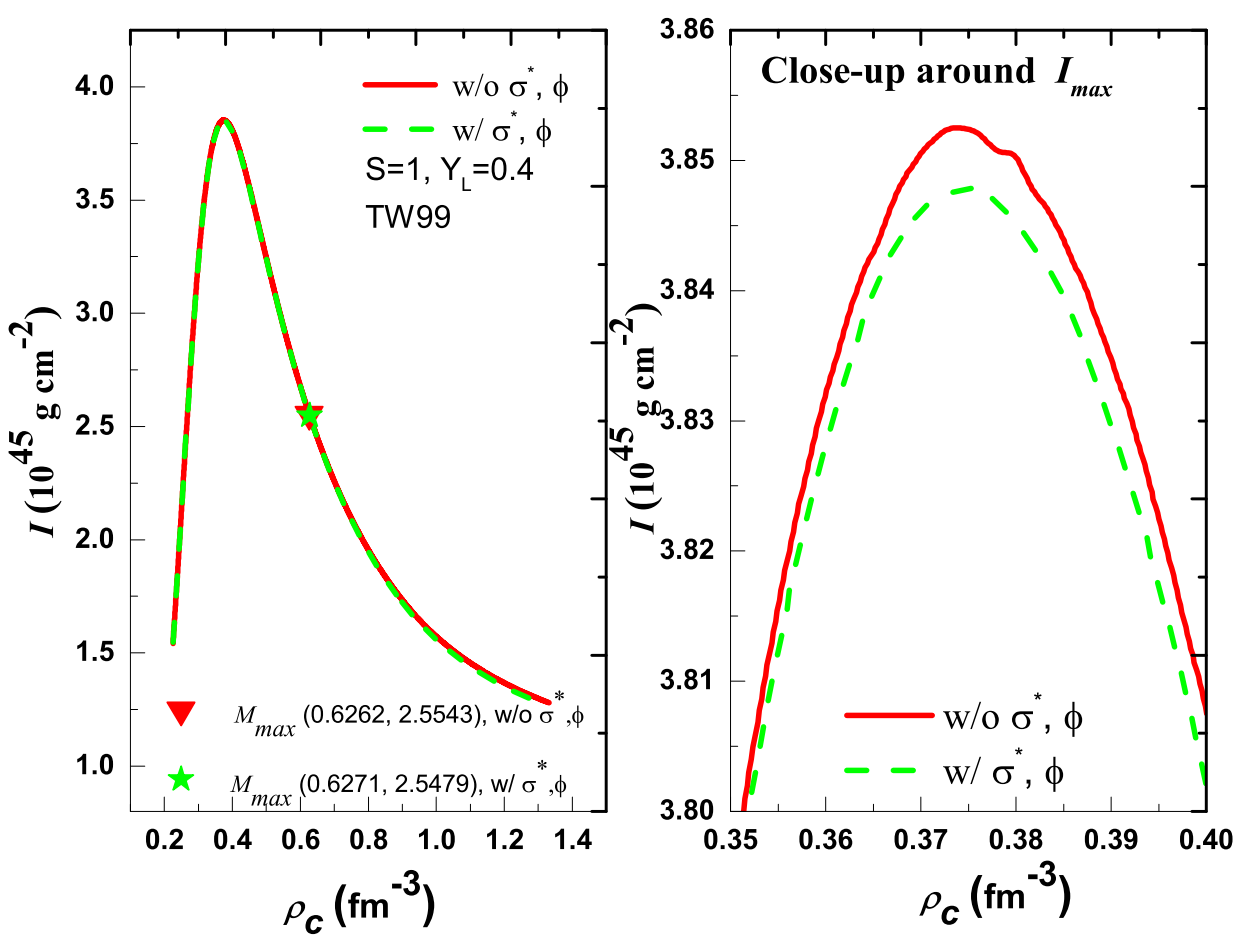

**Fig. 4.** Moment of inertia as a function of central baryon density for PNSs with the TW99 parametrization, entropy per baryon $S$=1 , and lepton fraction $Y_L$=0.4 . The left panel presents the overall behavior of the moment of inertia versus central baryon density, while the right panel provides a zoom-in view near the maximum moment of inertia. The red solid curve represents the case without strange mesons ( $\sigma^*$ and $\phi$ ), and the green dashed curve represents the case with strange mesons included. The red inverted triangle and the green five-pointed star mark the positions of the maximum moment of inertia ( $I_{\max}$ ) for the cases without and with strange mesons, respectively.

In the left panel, both curves exhibit a similar qualitative trend: the moment of inertia increases rapidly with density at low densities, then gradually slows down and reaches a maximum before decreasing slightly at higher densities. This non-monotonic behavior arises from the competition between mass growth and radius contraction. At low densities, the increase in mass dominates, leading to a rise in $I \sim MR^2$ . As the density increases further, the stellar radius begins to shrink significantly, eventually suppressing the growth of $I$ and causing it to turn over at $I_{\max}$.

The right panel reveals a clear shift in the $I_{\max}$ position when strange mesons are included. Specifically, the maximum moment of inertia occurs at a slightly higher central density in the presence of $\sigma^*$ and $\phi$ mesons. The corresponding $I_{\max}$ value is marginally lower when strange mesons are included. This shift reflects the softening of the EoS induced by the attractive interactions of strange mesons, which reduces the pressure support and leads to a more compact stellar configuration for a given central density. Consequently, the density at which the radius contraction starts to dominate the behavior of $I$ shifts toward higher densities.

The fact that $I_{\max}$ occurs at a significantly lower density than the maximum mass (as previously shown in Fig. 1) is also evident in this figure. The density separation between the $I_{\max}$ position and the $M_{\max}$ position is a generic feature of NS structure, originating from the different sensitivities of mass and moment of inertia to the density profile. The zoom-in panel clearly demonstrates that the inclusion of strange mesons does not alter this qualitative behavior but rather induces a systematic shift in the peak position and amplitude.

Quantitatively, the right panel shows that the relative change in the peak value of $I$ due to strange mesons is small (on the order of 0.1% ), while the density shift is also modest. These small variations indicate that the global rotational properties of massive PNSs are only moderately affected by the presence of strange mesons. Nevertheless, the systematic nature of the shift—consistently toward higher densities and lower $I_{\max}$ values—suggests that strange mesons play a non-negligible role in fine-tuning the rotational characteristics of massive PNSs, particularly near the threshold density where hyperons and strange mesons become active.

Overall, Fig. 4 provides a clear visual demonstration of the influence of strange mesons on the moment of inertia, confirming that their primary effect is to shift the $I_{\max}$ position toward higher densities while slightly reducing its magnitude. This observation is consistent with the softening of the EoS discussed in Figures 2 and 3, and it establishes the foundation for quantifying the relative changes in $I$ presented in the following figures.

Figure 5 quantifies the shifts in the maximum moment of inertia ( $I_{\max}$) and its corresponding central density ( $\rho_c^{I_{\max}}$ ) induced by strange mesons. The vertical and horizontal dashed lines mark the changes in $I_{\max}$ and $\rho_c^{I_{\max}}$ , respectively. The inclusion of $\sigma^*$ and $\phi$ mesons shifts $\rho_c^{I_{\max}}$ to higher densities while slightly reducing $I_{\max}$. This behavior reflects the softening of the EoS due to the attractive interactions of strange mesons, which leads to a more compact configuration and shifts the balance between mass growth and radius contraction toward higher densities. The observed shifts are systematic but modest, indicating that strange mesons have a measurable, albeit moderate, influence on the rotational properties of massive PNSs near the threshold density where exotic degrees of freedom become active.

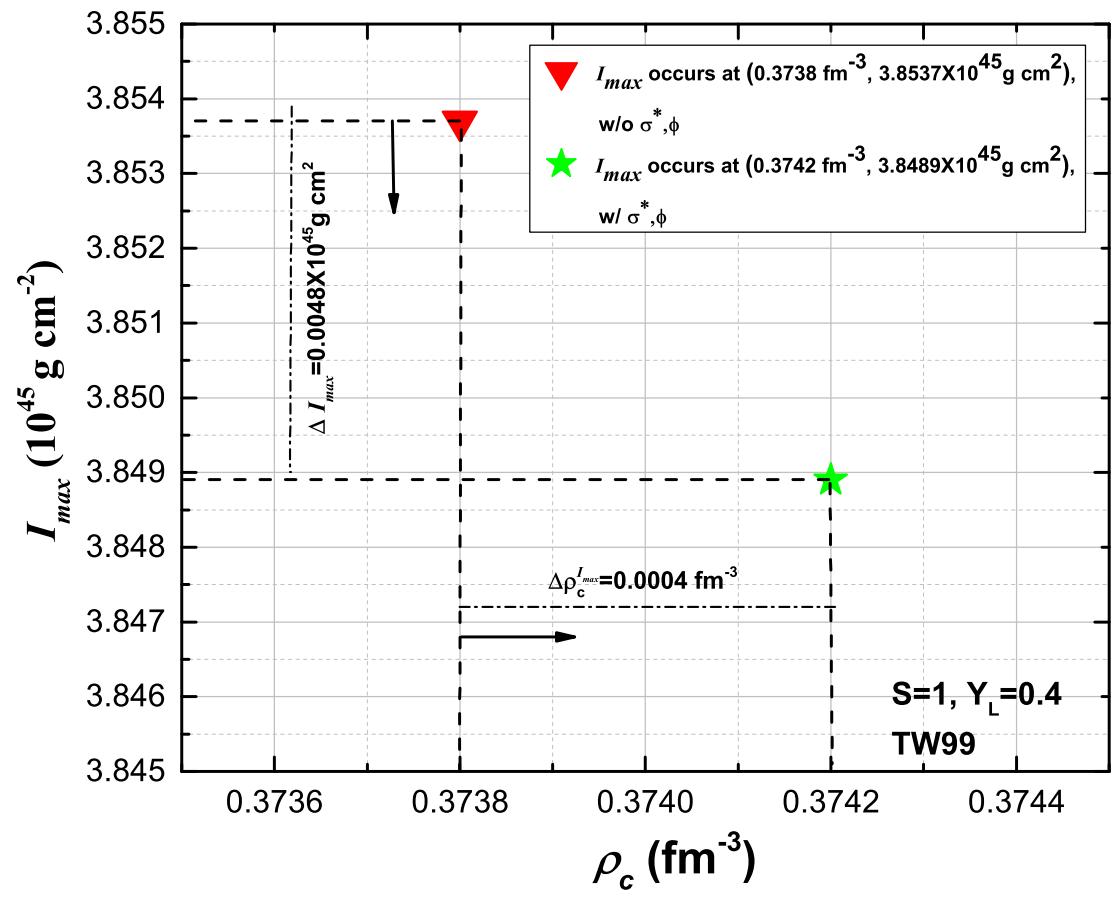

**Fig. 5.** Shift in the maximum moment of inertia and the corresponding central baryon density induced by the inclusion of strange mesons ($\sigma^*$ and $\phi$). The results are obtained with the TW99 parametrization, entropy per baryon $S$=1 , and lepton fraction $Y_L$=0.4 . The horizontal dashed line indicates the variation of the maximum moment of inertia, while the vertical dashed line indicates the variation of the central baryon density at which the maximum occurs.

Figure 6 presents the relative change rate of the moment of inertia, defined as

$$\delta I = \frac{(I_{with} - I_{without})}{I_{without}} \times 100\%, \tag{19}$$

as a function of gravitational mass for PNSs with the TW99 parametrization, $S$=1 , and $Y_L$=0.4 . For masses below approximately 2.1 $\mathrm{M}_\odot$ , $\delta I$ remains nearly zero, indicating that strange mesons have negligible influence on the moment of inertia in this mass range. As the mass exceeds 2.1 $\mathrm{M}_\odot$, $\delta I$ begins to decrease monotonically, reaching about -0.6% at 2.7 $\mathrm{M}_\odot$ . This trend reflects the density-dependent activation of strange mesons: only in the most massive PNSs, where the central density is sufficiently high, do $\sigma^*$ and $\phi$ mesons become abundant enough to soften the EoS and reduce the stellar radius, thereby suppressing the moment of inertia. The gradual decline of $\delta I$ with increasing mass suggests that the strangeness effect scales with the compactness of the star, becoming more pronounced as the star approaches its maximum mass. Overall, the mass threshold of $\sim$ 2.1 $\mathrm{M}_\odot$ marks the onset of significant strangeness contributions to the rotational properties of massive PNSs.

## 6 Summary

In this work, we have systematically investigated the effects of strange mesons ( $\sigma^*$ and $\phi$ ) on the structural and rotational properties of massive PNSs within the RMF framework. The calculations were performed with the TW99 nucleonic parametrization, fixed entropy per baryon $S$=1 , and lepton fraction $Y_L$=0.4 , conditions that are relevant for the early evolutionary stage of NSs.

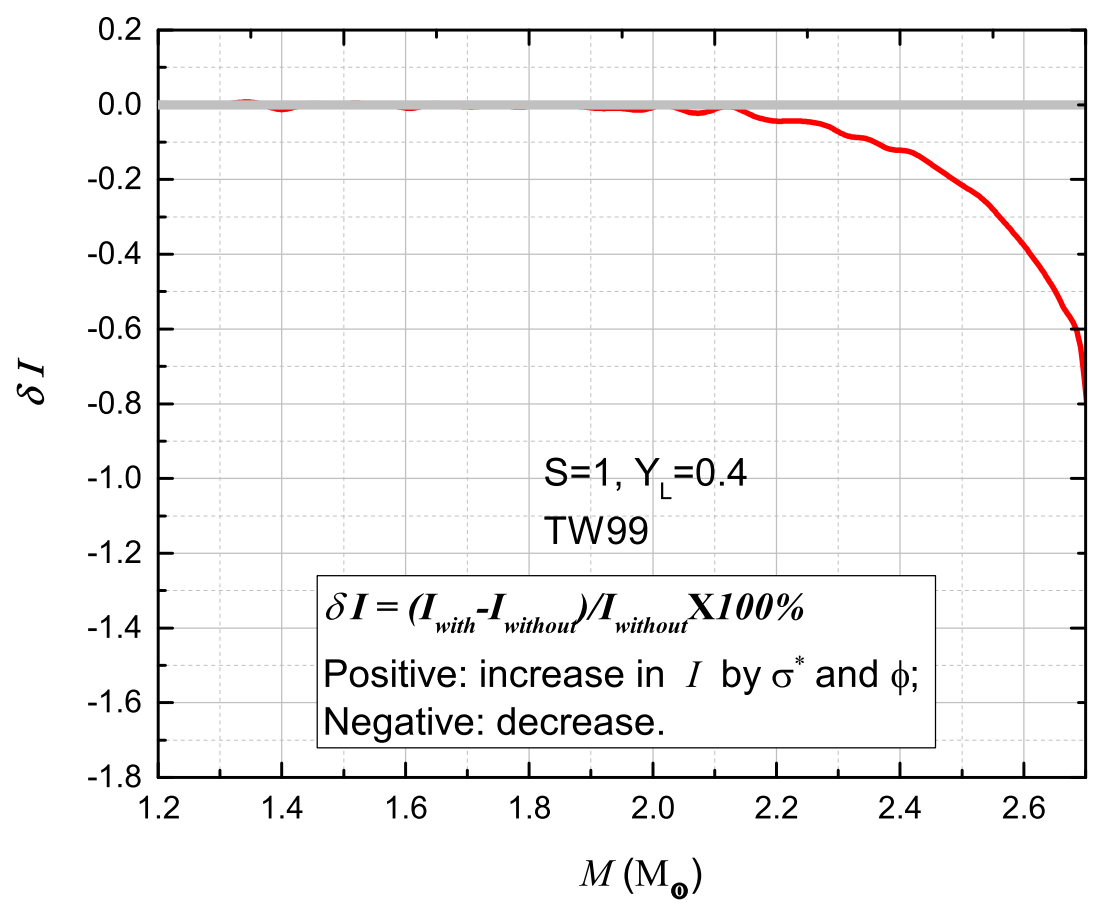

**Fig. 6.** Relative change rate of the moment of inertia as a function of gravitational mass for PNSs with the TW99 parametrization, entropy per baryon $S$=1 , and lepton fraction $Y_L$=0.4 . The relative change rate is defined as $\delta I = (I_{with} - I_{without})/I_{without} \times 100\%$ , where $I_{with}$ and $I_{without}$ denote the moments of inertia with and without strange mesons ( $\sigma^*$ and $\phi$ ), respectively. Positive values indicate that the inclusion of strange mesons increases the moment of inertia, while negative values indicate a decrease.

Among eight nucleonic parametrizations examined, TW99 was selected as it yields the largest maximum gravitational mass, making it suitable for describing massive PNSs. We found that the inclusion of strange mesons softens the EoS at high densities, leading to a systematic reduction of both the maximum mass and the corresponding radius. The softening effect is most pronounced in the density regime where hyperons and strange mesons become abundant.

A key finding is the distinct behavior of the moment of inertia. Unlike the gravitational mass, which reaches its maximum at the highest central density, the moment of inertia attains its peak at a significantly lower density. The inclusion of strange mesons shifts this peak toward slightly higher densities while reducing its magnitude. The relative change in the moment of inertia remains negligible for PNSs with masses below 2.1 $M_\odot$, but becomes increasingly negative for more massive stars, reaching about -0.6% at 2.7 $M_\odot$. This mass-dependent behavior indicates that strange mesons primarily affect the rotational properties of the most massive PNSs.

Overall, our results demonstrate that while strange mesons have only moderate effects on the absolute values of the moment of inertia, they induce systematic and physically meaningful shifts in the peak position and magnitude of $I$. These findings provide useful constraints for understanding the rotational evolution of massive PNSs and may have implications for the gravitational-wave signatures expected from future supernova or NS merger events. Accurate modeling of the strange meson sector remains essential for reliable predictions of PNS properties near the mass limit.

## Acknowledgments

This work was supported by the Natural Science Foundation of China (Grant No. 12465023).